\documentclass[
reprint,
amsmath,amssymb,
aps,
pra,
]{revtex4-2}

\usepackage{graphicx, color}
\usepackage{dcolumn}
\usepackage{bm}
\usepackage{here}
\usepackage[whole]{bxcjkjatype}

\begin{document}

\title{Direct measurement of the $5s5p\,{}^1P_1 \to 5s4d\,{}^1D_2$ decay rate in strontium}

\author{Naohiro Okamoto}
\author{Takatoshi Aoki}
\author{Yoshio Torii}
\email{ytorii@phys.c.u-tokyo.ac.jp}

\affiliation{Institute of Physics, The University of Tokyo, 3-8-1 Komaba, Meguro-ku, Tokyo 153-8902, Japan}
\date{\today}

\begin{abstract}
We report the first direct experimental determination of the branching ratio of the 
$5s4d\,{}^1D_2 \to 5s5p\,{}^3P_2$ transition and the decay rate of the 
$5s5p\,{}^1P_1 \to 5s4d\,{}^1D_2$ transition in neutral strontium.
For more than four decades, these quantities lacked an experimental determination independent of theoretical input.
We measure the branching ratio to be $0.177(4)$, significantly lower than the widely cited theoretical value of $0.322$ [C. W. Bauschlicher Jr. {\it et al.}, J. Phys. B \textbf{18}, 1523 (1985)]. 
We also determine the decay rate to be $5.3(5)\times10^3\,\mathrm{s^{-1}}$, consistent with the value reported by Hunter [L. R. Hunter {\it et al.}, Phys. Rev. Lett. \textbf{56}, 823 (1986)] but substantially lower than the recent theoretical prediction of $9.25(40)\times10^3\,\mathrm{s^{-1}}$ [A. Cooper {\it et al.}, Phys. Rev. X \textbf{8}, 041055 (2018)].
These measurements provide an experimental benchmark for quantitative modeling of loss processes in laser cooling and single-atom fluorescence detection in optical tweezers with Sr.


\end{abstract}

\maketitle

Alkaline-earth-metal (-like) atoms possess long-lived metastable states and ultra-narrow optical transitions, enabling a broad range of applications, including precision metrology~\cite{S.L.Campbell2017, E.Oelker2019, T.L.Nicholson2015, W.F.McGrew2018, S.M.Brewer2019, T.Bothwell2019, N.Nemitz2016, BACONcolab2021}, tests of special relativity~\cite{P.Delva2017}, measurements of gravitational redshift~\cite{M.Takamoto2020, T.Bothwell2022, X.Zheng2023}, quantum simulation~\cite{S.Kolkowits2017}, quantum information~\cite{A.Daley2008, N.Schine2022, R.Tao2025}, gravitational wave detection~\cite{S.Kolkowits2016, M.Abe2021}, and search for dark matter~\cite{M.Abe2021, T.Kobayashi2022}.
Among these, strontium-based optical lattice clocks have been extensively investigated as leading candidates for a redefinition of the second~\cite{N.Dimarcq2024}.

\begin{figure}
	\begin{center}
		\includegraphics[width=86mm]{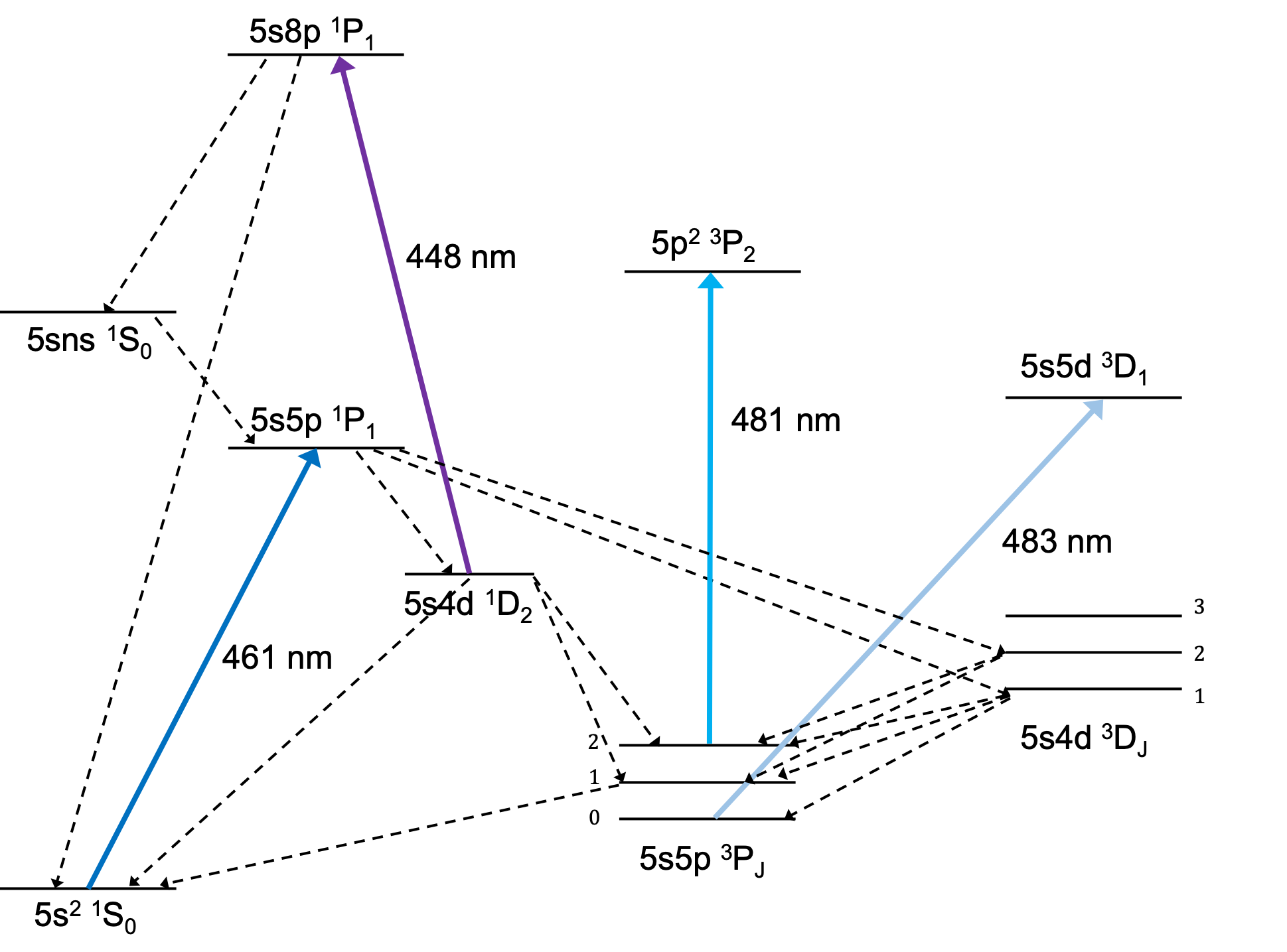}
		\caption{Energy level diagram of Sr relevant to this study.}
		\label{fig:energy_level}
	\end{center}
\end{figure}

In a standard magneto-optical trap (MOT) for strontium (Sr) atoms, the $5s^2\,{}^1S_0 - 5s5p\,{}^1P_1$ transition at $461\,\mathrm{nm}$ is typically employed (Fig.~\ref{fig:energy_level}). 
This transition is not completely closed, as a small fraction of atoms in the $5s5p\,{}^1P_1$ state decays to the $5s4d\,{}^1D_2$ state.
Subsequently, atoms in the $5s4d\,{}^1D_2$ state decay either to the $5s5p\,{}^3P_2$ state or to the $5s5p\,{}^3P_1$ state via spin-forbidden transitions. 
Atoms decaying to the $5s5p\,{}^3P_1$ state quickly decay to the ground state and return to the cooling cycle. 
In contrast, atoms decaying to the metastable $5s5p\,{}^3P_2$ state, with a long lifetime of approximately $10^3\,\mathrm{s}$~\cite{A.Derevianko2001, M.Yasuda2004}, are lost from the cooling cycle.

The effective loss rate per ${}^1P_1$ atom is given by the product of two physical quantities: the decay rate of the $5s5p\,{}^1P_1 \to 5s4d\,{}^1D_2$ transition ($A_{{}^1P_1\to{}^1D_2}$) and the branching ratio of the $5s4d\,{}^1D_2 \to 5s5p\,{}^3P_2$ transition ($B_{{}^1D_2\to{}^3P_2}$). 
The decay rate was indirectly measured by Hunter {\it et al.}~\cite{L.R.Hunter1986} via Stark-induced E1 transition experiments to be $3.9(1.5)\times10^3\,\mathrm{s^{-1}}$. 
Given that the decay rate for the $5s^2\,{}^1S_0 - 5s5p\,{}^1P_1$ transition is $1.900(1)\times10^8\,\mathrm{s^{-1}}$~\cite{M.Yasuda2006}, the branching ratio of the $5s5p\,{}^1P_1 \to 5s4d\,{}^1D_2$ transition is approximately $1:50\,000$.
All possible decay rates from the $5s4d\,{}^1D_2$ state were theoretically calculated by Bauschlicher {\it et al.}~\cite{C.W.Bauschlicher1985}, yielding a branching ratio $B_{{}^1D_2\to{}^3P_2}$ of $0.322$ ($\sim$1/3). 
We note that spin-forbidden transitions arise from singlet–triplet mixing induced by spin–orbit interaction, and that the apparent proximity of the branching ratio $B_{{}^1D_2\to{}^3P_2}$ to 1/3 is merely accidental rather than a result of selection rules. Consequently, the effective loss rate per ${}^1P_1$ atom is $A_{{}^1P_1\to{}^1D_2}B_{{}^1D_2\to{}^3P_2} = 1.3(5)\times 10^3\,\mathrm{s^{-1}}$ ($1:150\,000$), which has been a widely acknowledged value~\cite{X.Xu2003} and is consistent with our recent measurement of $A_{{}^1P_1 \to {}^1D_2}B_{{}^1D_2 \to {}^3P_2} = 9.3(9)\times10^2\,\mathrm{s^{-1}}$~\cite{N.Okamoto2025}.

In contrast, various theoretical calculations have proposed different values for $A_{{}^1P_1\to{}^1D_2}$: Bauschlicher {\it et al.} reported $6.1(2.2)\times10^3\,\mathrm{s^{-1}}$~\cite{C.W.Bauschlicher1985}, Werij {\it et al.} obtained $1.7\times10^4\,\mathrm{s^{-1}}$~\cite{H.G.C.Werij1992}, Porsev {\it et al.} calculated $9.1(3.8)\times10^3\,\mathrm{s^{-1}}$~\cite{S.G.Porsev2001}, and Cooper {\it et al.} reported $9.25(40)\times10^3\,\mathrm{s^{-1}}$~\cite{A.Cooper2018}.
Notably, Cooper's value corresponds to a branching ratio of $1:20,000$, approximately 2.5 times larger than the widely cited experimental value of $1:50\,000$.
Cooper's value is consistent with experimental results derived from fluorescence detection of single atoms trapped in optical tweezers, which provides an upper bound on the branching ratio in the range from $1:24(4)\times10^3$ to $1:17(3)\times10^3$~\cite{A.Cooper2018}. 
As a result, two distinct sets of values for the decay rate and branching ratio for the $5s5p\,{}^1P_1 \to 5s4d\,{}^1D_2$ transition currently coexist in the literature~\cite{J.Samland2024}.

The experimental value of $A_{{}^1P_1\to{}^1D_2}$ reported by Hunter {\it et al.}~\cite{L.R.Hunter1986} relies on the theoretical value of the $5s4d\,{}^1D_2 - 5s^2\,{}^1S_0$ electric quadrupole transition rate calculated by Bauschlicher {\it et al.}~\cite{C.W.Bauschlicher1985}.
In this sense, Hunter's value is not purely experimental.
To date, no direct measurement of $A_{{}^1P_1\to{}^1D_2}$ independent of theoretical calculations has been conducted.

In this work, we investigate the decay process $5s5p\,{}^1P_1 \to 5s4d\,{}^1D_2 \to 5s5p\,{}^3P_2$ in a MOT of $\mathrm{{}^{88}Sr}$ atoms by irradiating the trapped atoms with laser light resonant with the $448\,\mathrm{nm}$ ($5s4d\,{}^1D_2 - 5s8p\,{}^1P_1$) transition and observing the transient response of atom fluorescence.
First, we measure, for the first time, the branching ratio $B_{{}^1D_2\to{}^3P_2} = 0.177(4)$, which significantly deviates from Bauschlicher's theoretical value.
Next, we experimentally determine, without relying on theoretical calculations, the decay rate $A_{{}^1P_1\to{}^1D_2} = 5.3(5)\times10^3\,\mathrm{s^{-1}}$. 
This result is consistent with Hunter's experimental value, but significantly lower than Cooper's theoretical value. 
The decay rate $A_{{}^1P_1\to{}^1D_2}$ obtained in our work is crucial not only for determining the loss rate in laser cooling of Sr but also for evaluating the survival probability in single-atom fluorescence detection in optical tweezers~\cite{J.Samland2024, A.Cooper2018}. 
Moreover, our findings call for a reevaluation of the previous theoretical frameworks used to calculate the decay rates of the $5s5p\,{}^1P_1 \to 5s4d\,{}^1D_2$ and $5s4d\,{}^1D_2 \to 5s5p\,{}^3P_{1,2}$ transitions.

Our experiment is based on observing the transient population dynamics of atoms in the ${}^1D_2$ state via the fluorescence signal of a MOT operating on the $461\,\mathrm{nm}$ transition.
The experimental apparatus is similar to that described in Ref.~\cite{N.Okamoto2025, N.Okamoto2025RSI}. 
Figure~\ref{fig:energy_level} shows the relevant energy levels of Sr for this study. 
To fully close the cooling cycle driven by the $461\,\mathrm{nm}$ ($5s^2\,{}^1S_0 - 5s5p\,{}^1P_1$) transition, we employ laser light at $481\,\mathrm{nm}$ ($5s5p\,{}^3P_2 - 5p^2\,{}^3P_2$) transition for ${}^3P_2$ repumping and laser light at $483\,\mathrm{nm}$ ($5s5p\,{}^3P_0 - 5s5d\,{}^3D_1$) transition for ${}^3P_0$ repumping~\cite{N.Okamoto2024}. 
In addition, laser light resonant with the $448\,\mathrm{nm}$ ($5s4d\,{}^1D_2 - 5s8p\,{}^1P_1$) transition is applied to optically pump atoms in the $5s4d\,{}^1D_2$ state back to the ground state~\cite{J.Samland2024}.
Since the lifetime of the upper state of the $448\,\mathrm{nm}$ transition ($\sim 30\,\mathrm{ns}$~\cite{J.Samland2024}) is much shorter than that of the lower state ($\sim 400\,\mathrm{\mu s}$~\cite{D.Husain1988}), the population in the ${}^1D_2$ state remains negligibly small while the $448\,\mathrm{nm}$ pumping light is on.

The MOT is loaded from a thermal atomic beam emitted from a Sr oven operated at $340\,^\circ\mathrm{C}$~\cite{N.Okamoto2025RSI}. 
The trapped atom number is $N \approx 4\times10^{6}$ and the background pressure is $\sim 4\times10^{-10}\,\mathrm{Torr}$. 
The one-body loss rate due to background-gas collisions is approximately $0.07\,\mathrm{s^{-1}}$, while the effective one-body loss rate due to two-body collisions is about $0.4\,\mathrm{s^{-1}}$. These rates are more than three orders of magnitude smaller than the natural decay rate 
of the ${}^1D_2$ state ($1/400\,\mathrm{\mu s} \approx 2.5\times10^{3}\,\mathrm{s^{-1}}$). 
Therefore, collisional loss processes are negligible on the time scale of the ${}^1D_2$ population dynamics relevant to the present analysis.

The rate equation describing the transient response of the number of atoms in the ${}^1D_2$ state after switching off the $448\,\mathrm{nm}$ light is then given by
\begin{equation}
    \frac{dN_{{}^1D_2}(t)}{dt} = fA_{{}^1P_1 \to {}^1D_2}N(t) - \gamma_{{}^1D_2} N_{{}^1D_2}(t), \label{eq:1D2_rate1}
\end{equation}
where $\gamma_{{}^1D_2} = A_{{}^1D_2 \to {}^3P_2} + A_{{}^1D_2 \to {}^3P_1} + A_{{}^1D_2 \to {}^1S_0}$ is the total natural decay rate from the ${}^1D_2$ state, $f$ is the fraction of atoms in the excited state ($5s5p\,{}^1P_1$) of the $461\,\mathrm{nm}$ cooling transition, and $A_X$ denotes the decay rate of transition $X$. 
Here, $N(t)$ is the total number of atoms in the $461\,\mathrm{nm}$ cooling cycle, and $N_{{}^1D_2}(t)$ is the number of atoms in the ${}^1D_2$ state.

If we neglect the population in the $5s5p\,{}^3P_1$ state, which has a short lifetime of $21.28(3)\,\mathrm{\mu s}$~\cite{T.L.Nicholson2015}, the following atom number conservation relation holds:
\begin{equation}
    N(t) + N_{{}^1D_2}(t) = N_0, \label{eq:number_conservation}
\end{equation}
where $N_0$ is the trapped atom number before the $448\,\mathrm{nm}$ light is switched off. 
Combining Eqs.~\eqref{eq:1D2_rate1} and \eqref{eq:number_conservation}, we obtain
\begin{equation}
    \frac{dN_{{}^1D_2}(t)}{dt} = \alpha \gamma_{{}^1D_2}N_0 - \gamma N_{{}^1D_2}(t), \label{eq:1D2_rate3}
\end{equation}
where $\alpha = \frac{fA_{{}^1P_1 \to {}^1D_2}}{\gamma_{{}^1D_2}}$ and $\gamma = (1+\alpha)\gamma_{{}^1D_2}$.
When the $448\,\mathrm{nm}$ light is turned off at $t=0$, the solution to Eq.~\eqref{eq:1D2_rate3} is
\begin{equation}
    N_{{}^1D_2}(t) = \frac{\alpha}{1+\alpha} N_0 \left[1-\exp(-\gamma t)\right]. \label{eq:1D2_solution}
\end{equation}
Substituting Eq.~\eqref{eq:1D2_solution} into Eq.~\eqref{eq:number_conservation} gives
\begin{equation}
    N(t) = N_s\left[1+\alpha\exp(-\gamma t)\right], \label{eq:MOT_steady}
\end{equation}
where $N_s = N_0/(1+\alpha)$ is the steady-state atom number after the $448\,\mathrm{nm}$ light is switched off. 
Thus, the trapped atom number $N(t)$ observed via fluorescence decreases exponentially toward $N_s$ with a decay rate $\gamma$. 
From the ratio $N_s/N_0$, $\alpha$ can be determined, and in combination with $\gamma$, both $\gamma_{{}^1D_2}$ and $fA_{{}^1P_1 \to {}^1D_2}$ can be extracted. 
For a rigorous treatment including the population in the $5s5p\,{}^3P_1$ state, which is used for the actual derivation of $\gamma_{{}^1D_2}$ and $fA_{{}^1P_1 \to {}^1D_2}$, see Appendix~\ref{append:3P1}.

As described later, measurement of the MOT loss rate allows us to determine the product $fA_{{}^1P_1 \to {}^1D_2}B_{{}^1D_2 \to {}^3P_2}$, where $B_{{}^1D_2 \to {}^3P_2} = A_{{}^1D_2 \to {}^3P_2}/\gamma_{{}^1D_2}$ denotes the branching ratio from ${}^1D_2$ to ${}^3P_2$. 
Once the value of $fA_{{}^1P_1 \to {}^1D_2}$ is determined by the method described above, $B_{{}^1D_2 \to {}^3P_2}$ can be extracted. 
Moreover, based on our previous study~\cite{N.Okamoto2025}, where $A_{{}^1P_1 \to {}^1D_2}B_{{}^1D_2 \to {}^3P_2} = 9.3(9)\times10^2\,\mathrm{s^{-1}}$, the decay rate $A_{{}^1P_1 \to {}^1D_2}$ can also be obtained.

\begin{figure}[ht]
	\begin{center}
		\includegraphics[width=86mm]{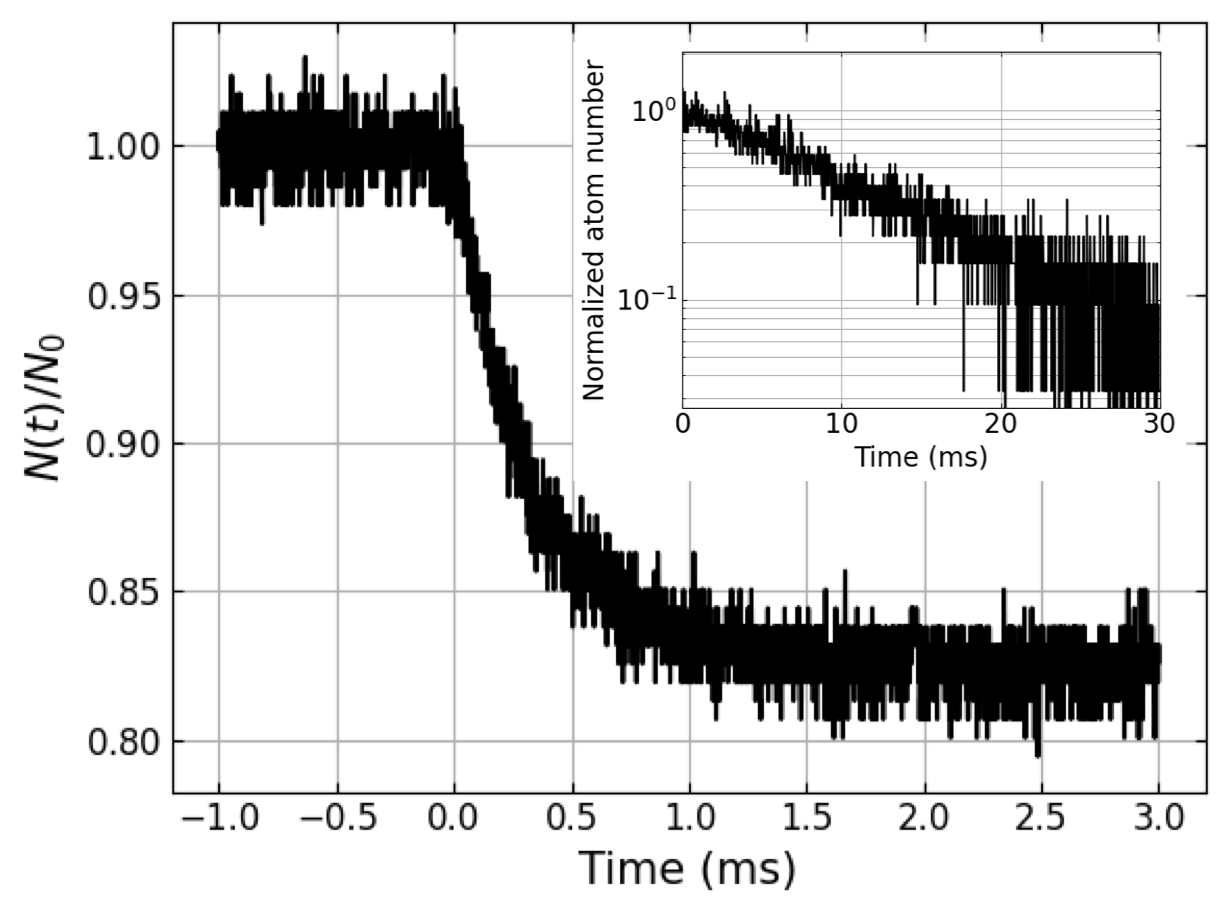}
		\caption{Change in the trapped atom number after switching off the $448\,\mathrm{nm}$ light, with the $461\,\mathrm{nm}$, $481\,\mathrm{nm}$, and $483\,\mathrm{nm}$ lights kept on. The data represent the average of 64 experimental traces. The inset shows the change in the trapped atom number after switching off the $481\,\mathrm{nm}$ light while the $461\,\mathrm{nm}$ and $483\,\mathrm{nm}$ lights remain on. The detuning of the MOT beams is set to $-36\,\mathrm{MHz}$.}
		\label{fig:448nmoff}
	\end{center}
\end{figure}

\begin{figure}[hhhhh]
	\begin{center}
		\includegraphics[width=86mm]{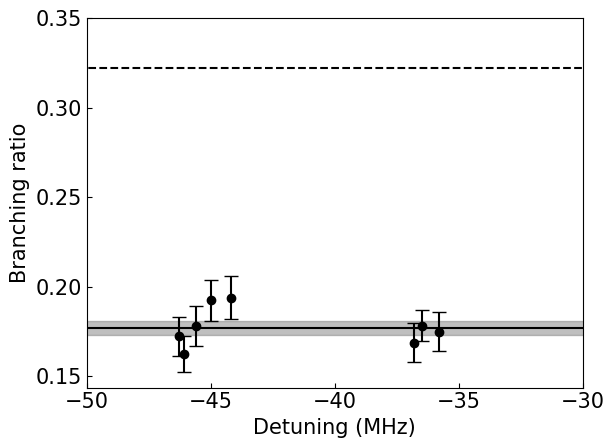}
		\caption{Extracted branching ratios for different trapping light detunings. The error bars represent the uncertainties derived from the fits to the decay curves as shown in Fig.~\ref{fig:448nmoff}. The solid line indicates the weighted average of the data, and the gray band represents the overall $\pm 1\sigma$ statistical uncertainty. For reference, the dashed line shows the theoretical value ($0.322$) reported by Bauschlicher {\it et al.}~\cite{C.W.Bauschlicher1985}. The reduced $\chi^2$ value for the combined data is 0.96, corresponding to a $P$ value of 0.46, indicating that the branching ratio shows no statistically significant dependence on the detuning.}
		\label{fig:branching_ratio}
	\end{center}
\end{figure}

Figure~\ref{fig:448nmoff} shows the decrease in the trapped atom number when the $448\,\mathrm{nm}$ light is switched off, starting from the steady state with all the $461\,\mathrm{nm}$, $481\,\mathrm{nm}$, $483\,\mathrm{nm}$, and $448\,\mathrm{nm}$ lights on. 
By fitting the data using Eq.~\eqref{eq:MOT_steady}, we determine $\gamma_{{}^1D_2} = 2.37(1)\times10^3\,\mathrm{s^{-1}}$, which agrees well with the experimental result $2.43(6)\times10^3\,\mathrm{s^{-1}}$ reported in Ref.~\cite{D.Husain1988}.

The inset of Fig.~\ref{fig:448nmoff} shows the decrease in the trapped atom number when the $481\,\mathrm{nm}$ light is switched off, while keeping the $461\,\mathrm{nm}$ and $483\,\mathrm{nm}$ lights on.
In this case, the rate $L$ of the exponential decay can be expressed as
\begin{align}
    L &= f\big(A_{{}^1P_1\to{}^1D_2}B_{{}^1D_2\to{}^3P_2} + A_{{}^1P_1\to{}^3D_2}B_{{}^3D_2\to{}^3P_2} \notag \\
    &\quad \quad + A_{{}^1P_1\to{}^3D_1}B_{{}^3D_1\to{}^3P_2}\big).
\end{align}

From our previous study~\cite{N.Okamoto2025}, 
the relative contributions of the decay paths are $91.4(3)\%$, $4.6(2)\%$, and $0.120(5)\%$ for $A_{{}^1P_1\to{}^1D_2}B_{{}^1D_2\to{}^3P_2}$, $A_{{}^1P_1\to{}^3D_2}B_{{}^3D_2\to{}^3P_2}$, and $A_{{}^1P_1\to{}^3D_1}B_{{}^3D_1\to{}^3P_2}$, respectively.
Using these values, we determine $fA_{{}^1P_1 \to {}^1D_2}B_{{}^1D_2 \to {}^3P_2}$ and extract the branching ratio $B_{{}^1D_2\to{}^3P_2}$.
Figure~\ref{fig:branching_ratio} shows the branching ratio $B_{{}^1D_2 \to {}^3P_2}$ for various trapping light detunings.
By taking the weighted average of the data, we obtain $B_{{}^1D_2\to{}^3P_2}=0.177(4)$, which significantly deviates from the widely cited value $\sim 1/3$~\cite{C.W.Bauschlicher1985}.

\begin{figure}
	\begin{center}
		\includegraphics[width=86mm]{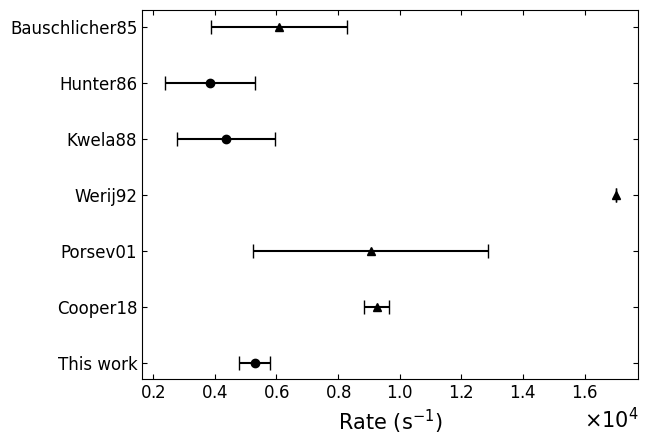}
		\caption{Comparison of the literature values and the present measurement for the $5s5p\,{}^1P_1 - 5s4d\,{}^1D_2$ transition rate~\cite{C.W.Bauschlicher1985, L.R.Hunter1986, J.Kwela1988, H.G.C.Werij1992, S.G.Porsev2001, A.Cooper2018}. Circular markers indicate experimental results, and triangular markers indicate theoretical values.}
		\label{fig:rate_comparison}
	\end{center}
\end{figure}

Using $A_{{}^1P_1\to{}^1D_2}B_{{}^1D_2\to{}^3P_2}$ obtained in our previous study~\cite{N.Okamoto2025} and the present result of $B_{{}^1D_2\to{}^3P_2}$, we determine $A_{{}^1P_1\to{}^1D_2} = 5.3(5)\times10^3\,\mathrm{s^{-1}}$.
A comparison of the literature values and the present result for $A_{{}^1P_1\to{}^1D_2}$ is shown in Fig.~\ref{fig:rate_comparison}.
The present result is consistent with Hunter's experimental value but significantly lower than Cooper's and Werij's theoretical values.

A summary of the results obtained in this work is presented in Table~\ref{table:results}. 
Here, the decay rate $A_{{}^1D_2\to{}^3P_1}$ is derived from $(1-B_{{}^1D_2\to{}^3P_2})\gamma_{{}^1D_2} - A_{{}^1D_2\to{}^1S_0}$, where the uncertainty comes from that in $A_{{}^1D_2\to{}^1S_0}$ ($\sim 100\,\mathrm{s^{-1}}$)~\cite{C.W.Bauschlicher1985}.
The branching ratio of the $5s5p\,{}^1P_1 \to 5s4d\,{}^1D_2$ transition is $1:36\,000\pm3\,000$, which lies between Hunter's value ($1:50\,000$) and Cooper's value ($1:20\,000$).

\begin{table}[tt]
    \caption{Measured decay rates and branching ratios.}
    \label{table:results}
    \centering
    \begin{tabular}{ccc}
    \hline \hline
    Transition  & Decay rate ($\mathrm{s^{-1}}$) & Branching ratio \\ \hline
    $5s4d\,{}^1D_2 - 5s5p\,{}^3P_2$ & $418(9)$ & 0.177(4) \\
    $5s4d\,{}^1D_2 - 5s5p\,{}^3P_1$ & $1.95(10)\times10^3$ & 0.82(4) \\
    $5s5p\,{}^1P_1 - 5s4d\,{}^1D_2$ & $5.3(5)\times10^3$ & $1:3.6(3)\times10^4$ \\
    \hline \hline
    \end{tabular}
\end{table}

In conclusion, we investigated the decay pathway $5s5p\,{}^1P_1 \to 5s4d\,{}^1D_2 \to 5s5p\,{}^3P_2$ to address a long-standing lack of direct experimental data that has persisted for more than four decades. 
We measured, for the first time, the branching ratio of the $5s4d\,{}^1D_2 \to 5s5p\,{}^3P_2$ transition to be $0.177(4)$, which deviates from Bauschlicher's theoretical value.
We also determined the $5s5p\,{}^1P_1 \to 5s4d\,{}^1D_2$ decay rate as $5.3(5)\times10^3\,\mathrm{s^{-1}}$ without relying on theoretical calculations. 
This value is consistent with Hunter's value but significantly lower than Cooper's and Werij's values. 
The present work addresses a long-standing lack of direct experimental data on this decay pathway in Sr that has persisted for more than four decades.
These results are critical for understanding atom losses in Sr laser cooling and for evaluating the survival probability in single-atom fluorescence detection. Furthermore, our findings raise concerns about the validity of the previous theoretical calculations of the $5s5p\,{}^1P_1 \to 5s4d\,{}^1D_2$ and $5s4d\,{}^1D_2 \to 5s5p\,{}^3P_{1,2}$ decay rates~\cite{C.W.Bauschlicher1985, H.G.C.Werij1992, S.G.Porsev2001, A.Cooper2018}. 

This work was supported by JSPS KAKENHI Grant Numbers 23K20849 and 22KJ1163.

\appendix
\section{Rate equations including the population in the $5s5p\,{}^3P_1$ state}
\label{append:3P1}

In the main text, we neglect the population in the ${}^3P_1$ state, which has a short lifetime of $A_{{}^3P_1 \to {}^1S_0}^{-1} = 21.28(3)\,\mathrm{\mu s}$~\cite{T.L.Nicholson2015} compared with the lifetime of the ${}^1D_2$ state ($\gamma_{{}^1D_2}^{-1}\sim 400\,\mathrm{\mu s}$). 
For the actual derivation of $\gamma_{{}^1D_2}$ and $fA_{{}^1P_1 \to {}^1D_2}$, we extend the rate equations in the main text by incorporating the population in the ${}^3P_1$ state.

Let $N_{{}^3P_1}(t)$ denote the number of atoms in the ${}^3P_1$ state. 
The atom number conservation equation and the rate equations can then be written as
\begin{align}
    N_0 &= N + N_{{}^1D_2} + N_{{}^3P_1}, \label{eq:append_num_conserve}\\
    \frac{dN_{{}^1D_2}}{dt} &= fA_{{}^1P_1 \to {}^1D_2}N - \gamma_{{}^1D_2}N_{{}^1D_2}, \label{eq:append_1D2_rate1}\\
    \frac{dN_{{}^3P_1}}{dt} &= \gamma_{{}^1D_2}N_{{}^1D_2} - A_{{}^3P_1 \to {}^1S_0}N_{{}^3P_1}, \label{eq:append_3P1_rate1}
\end{align}
where $N$ is the number of atoms in the $461\,\mathrm{nm}$ cooling cycle, $N_0$ is the trapped atom number before the $448\,\mathrm{nm}$ light is switched off, and $N_{{}^1D_2}$, $N_{{}^3P_1}$ are the number of atoms in the ${}^1D_2$, ${}^3P_1$ state, respectively.
Here, since $A_{{}^3P_1 \to {}^1S_0} \gg \gamma_{{}^1D_2}$, we can apply an adiabatic approximation $\frac{dN_{{}^3P_1}}{dt}=0$ to Eq.~\eqref{eq:append_3P1_rate1}, which leads to
\begin{align}
    N_{{}^3P_1} &= \theta N_{{}^1D_2}, \label{eq:append_num_3P1}\\
    \theta &= \frac{\gamma_{{}^1D_2}}{A_{{}^3P_1 \to {}^1S_0}}. \label{eq:append_theta}
\end{align}
Substituting Eqs.~\eqref{eq:append_num_conserve} and \eqref{eq:append_num_3P1} into Eq.~\eqref{eq:append_1D2_rate1}, we obtain
\begin{align}
    \frac{dN_{{}^1D_2}}{dt} &= \frac{\alpha}{1+\alpha+\theta\alpha}\frac{N_0}{\tau'} - \frac{N_{{}^1D_2}}{\tau'}, \label{eq:append_1D2_rate2}\\
    \tau' &= \frac{1}{\gamma_{{}^1D_2}(1+\alpha+\theta\alpha)}, \label{eq:append_tau}\\
    \alpha &= \frac{fA_{{}^1P_1\to{}^1D_2}}{\gamma_{{}^1D_2}}. \label{eq:append_alpha}
\end{align}
With the initial condition $N_{{}^1D_2}(0)=0$, Eq.~\eqref{eq:append_1D2_rate2} is solved as
\begin{equation}
    N_{{}^1D_2} = \frac{\alpha}{1+\alpha+\theta\alpha}N_0\left[1-\exp(-t/\tau')\right]. \label{eq:append_1D2_solution}
\end{equation}
Combining Eqs.~\eqref{eq:append_num_conserve}, \eqref{eq:append_num_3P1}, and \eqref{eq:append_1D2_solution}, the trapped atom number $N$ observed via fluorescence at $461\,\mathrm{nm}$ is given by
\begin{align}
    N &= N'_s\left[1+(1+\theta)\alpha\exp(-t/\tau')\right], \label{eq:append_num_MOT}\\
    N'_s &= \frac{N_0}{1+(1+\theta)\alpha}. \label{eq:append_steady_MOT}
\end{align}
Thus, the trapped atom number observed via $461\,\mathrm{nm}$ fluorescence decreases exponentially toward the steady-state value $N'_s$ with a time constant $\tau'$. 
From Eqs.~\eqref{eq:append_tau} and \eqref{eq:append_steady_MOT}, $\gamma_{{}^1D_2}$ is written with the measured $\tau'$ and $N'_s/N_0$ as
\begin{equation}
    \gamma_{{}^1D_2} = \frac{N'_s}{N_0}\frac{1}{\tau'}.
\end{equation}
Given the known value of $A_{{}^3P_1 \to {}^1S_0} = 4.699(7)\times10^4\,\mathrm{s^{-1}}$ and the value obtained in this work of $\gamma_{{}^1D_2}=2.37(1)\times10^3\,\mathrm{s^{-1}}$, $\theta$ is determined as 0.05, which leads to the determination of $\alpha$ and $fA_{{}^1P_1 \to {}^1D_2}$ using Eqs.~\eqref{eq:append_alpha} and \eqref{eq:append_steady_MOT}.

\bibliography{references.bib}

\end{document}